# Spin Rectification in Series and Parallel Networks of Ferromagnets

Md. Majibul Haque Babu*, Maxim Tsoi

*Department of Physics, The University of Texas at Austin, Austin, Texas 78712, USA*

Abstract

We present the experimental study of spin rectification in parallel and series networks of ferromagnetic conductors. The networks were composed of $Ni_{36}Fe_{64}$ elements (wires) which, when subject to ferromagnetic resonance (FMR), develop a small spin-rectification photovoltage. We found that the effect of spin rectification can be modeled as an increase in resistance of the ferromagnetic elements and closely follows the usual dependence of total network resistance in series and parallel electric circuits: the total spin-rectification photovoltage in ferromagnetic wires connected in series is equal to the algebraic sum of contributions from individual wires, while in the parallel connection one needs to add the reciprocals of the individual contributions. Our results support a simple way to control and manipulate the magnitude of spin-rectification photovoltage in multi-element spintronic devices.

## I. Introduction

The spin-rectification effect is a powerful tool to study magnetization dynamics in magnetic materials and spintronic devices [1–10]. For example, it provides a means to detect ferromagnetic resonance (FMR) electrically with high sensitivity. Here the effect originates from the dynamic coupling between an oscillating current and an oscillating resistance. When a microwave current $j_{rf}\cos(\omega t)$ flows through a metallic ferromagnet (F), e.g., an F wire of circular cross-section, it produces a circumferential Oersted field $h_{rf}\cos(\omega t)$ around the wire, which drives the wire magnetization into precession when the applied frequency $\omega$ matches the natural precession frequency of the magnetization. If the wire resistance depends on the orientation of its magnetization, e.g., via anisotropic magnetoresistance (AMR), the precession results in an oscillating resistance $\Delta R\cos(\omega t)$, where $\Delta R$ is the amplitude of resistance oscillations. The coupling between this oscillating resistance and the applied current generates a small dc photovoltage $V_\omega \sim j_{rf}\Delta R$, which can be routinely detected with a voltmeter.

We have previously shown [11-14] that a single F wire can be used as a test platform for studying the spin rectification. In this well-controlled geometry the rf current is confined to a thin surface layer of the wire [15], because the electromagnetic skin depth (~1 μm) at high frequencies ($\omega$ > 1 GHz) is much smaller than the wire diameter (~50 μm). Thus, the rf magnetic field $h_{rf}$ affects the wire magnetization only in a thin surface layer that makes our wire-like geometry equivalent to the case of a thin film in a parallel magnetic field – a well-studied FMR geometry. Our experiments with $Ni_{36}Fe_{64}$ wires demonstrated that FMR detection via spin-rectification effect is equivalent to the conventional microwave-absorption measurements [11-14] and can be used effectively to study the angular dependence of FMR [11], effects of sample shape on FMR [13], and also look into and distinguish between bulk and contact (local) rectification effects [12]. In our most recent work [14], we showed that the effect of spin rectification in FMR experiments can be modeled as an increase in resistance of the wire subject to FMR. This FMR-driven resistance increase was found to grow approximately linearly with the wire length, in direct analogy with the standard linear dependence of the resistance of an ordinary conductor on its length. The latter raises a natural question: if the spin-rectification response acts as an effective resistance, does it also combine like resistances of ordinary conductors in F-networks where several Fs are connected in series and/or in parallel?

*Contact author: majibulbabu@utexas.edu

In this paper we study the spin-rectification effect in series and parallel networks of $Ni_{36}Fe_{64}$ wires and address this question directly. We focus first on a series network, where several F elements are connected in series. We then reexamine the dependence of spin rectification on the length of a single $Ni_{36}Fe_{64}$ wire. This geometry is equivalent to the series network, and we extend our previous measurements [14] to extra-long wires of up to 6 mm. Finally, we study a parallel network, where up to six F elements are connected in parallel. We found that the scaling of spin rectification in networks of F elements is consistent with the classical circuit theory: the total spin-rectification photovoltage in series F-networks is equal to the algebraic sum of contributions from individual F elements, while for the parallel connection one needs to add the reciprocals of the individual contributions.

## II. Experimental Methods

In our experiments, we use sections of ferromagnetic $Ni_{36}Fe_{64}$ wire with $50\ \mu m$ diameter (Goodfellow FE02-WR-000100) as individual F-elements. This specific material and the wire diameter were chosen because our earlier experiments [11-14] showed a strong and reproducible spin-rectification signal in this simple geometry. It is also a very common ferromagnetic material, so we expect our results to be easily generalized for other ferromagnets. Note that an rf current applied to such a metallic wire is confined to a thin surface layer of the wire because the electromagnetic skin depth (~1 μm) is much smaller than the wire diameter [11, 15]. Thus, the rf magnetic field produced by the current affects the wire magnetization only in a thin surface layer that makes this wire-like geometry equivalent to the case of a thin film in a parallel magnetic field. We have previously confirmed [11, 12, 14] that FMR in such a wire follows the Kittel's resonance condition for a thin film $\omega/\gamma = \sqrt{H(H + 4\pi M_S)}$ [16], where $\omega$ is the rf frequency, $\gamma$ = 280 GHz/kOe is the gyromagnetic ratio, and $M_S$ = 957 emu/cm$^3$ is the saturation magnetization.

We have arranged our F-elements into two configurations – parallel and series networks. For the *parallel network*: $N$ identical $Ni_{36}Fe_{64}$ wires (~1.4 mm long) were soldered between two common electrodes. Fig. 1(a) shows a sample schematic for $N$ = 3. Samples with $N$ ranging from 1-6 were used in our experiments. Note that in such parallel networks the rf current applied to the sample is shared between $N$ wires approximately equally.

*Contact author: majibulbabu@utexas.edu

For the *series network*: one $Ni_{36}Fe_{64}$ wire (~1.4 mm long) was initially soldered between the electrodes. Then a conducting Ag paste was applied locally to the wire to create nonmagnetic (N) metal islands on the wire. Fig. 1(b) shows schematically a sample with two such islands. The F wire under the N islands is effectively short-circuited for rf signals by the island material (Ag); the applied rf current is confined to a thin (~ skin depth) surface layer in both $Ni_{36}Fe_{64}$ (F) and Ag (N) and, thus, forced to flow from F into N and back every time it crosses one of the islands. In Fig. 1(b) the two Ag islands divide the $Ni_{36}Fe_{64}$ wire into three F sections with lengths $L_1$, $L_2$, and $L_3$. The total magnetically active length is $L = L_1 + L_2 + L_3$ and the same amount of rf current flows through every section because there is only one single path for the current. We have prepared samples with up to four Ag islands; here four islands divide the wire into five F-sections. Fig. 1(c) shows a microscope image of $Ni_{36}Fe_{64}$ wire with one Ag-paste island (lighter color in the middle). We have also studied a single wire with a variable length. Here individual sections of the wire are not separated by N-elements but instead are "connected" directly to one another and an increased length of the wire corresponds to a "larger" number of elements connected in series. Our previous measurements on a 1.4-mm long $Ni_{36}Fe_{64}$ wire showed an increase of the spin-rectification voltage with the wire length [14]. Here we extend these measurements to extra-long wires of up to 6 mm. The maximum wire length in our experiments was limited by the magnet-gap size.

Fig. 1(d) shows a schematic of our experiment: a sample (parallel or series network; represented here by a single cylindrical wire) terminates a coaxial cable used to deliver rf and dc currents to the wire via a bias tee. An external dc magnetic field $H$ is applied along the wire/wires (in series/parallel networks). The rf current (of frequency $f$) applied to the wires produces a circumferential (rf) Oersted field $h_{rf}$ perpendicular to $H$. This rf magnetic field generates a torque on the wire's magnetization and drives it into precession (FMR) when the rf frequency $f=\omega/2\pi$ matches the resonance frequency. Black trace in Fig. 1(e) shows an example of FMR spectrum where the absorption of rf signal in the wire is measured as a function of the applied magnetic field $H$. The peak in absorption at about $H = \pm$ 0.5 kOe corresponds to FMR at $f$ =7 GHz. Red trace in Fig. 1(e) shows the spin-rectification FMR spectrum (measured simultaneously with the absorption spectrum) where the spin-rectification photovoltage $V_\omega$ is measured as a function of the applied magnetic field $H$. The peak in $V_\omega$ at about $H = \pm$ 0.5 kOe corresponds to FMR and displays spin rectification $V_\omega$ of ~0.5 mV. The black and red spectra in Fig. 1(e) are very similar and confirm that the electrical detection of FMR via the spin-rectification effect is equivalent to the conventional microwave-absorption measurements [11-14]. The insert in Fig. 1(e) shows that the voltage peak $V_\omega$ can be interpreted as a peak in the dc resistance $\Delta R_\omega=V_\omega/I_{dc}$, which exists only in the presence of $j_{rf}$, i.e., FMR produces an extra resistance of the wire. We now focus on this extra resistance produced in the series and parallel networks.

## III. Results and Discussion

First, we place our series network sample (schematically shown in Fig. 1(b)) – 1.4-mm-long F ($Ni_{36}Fe_{64}$) wire with four N-metal (Ag) islands – into the circuit of Fig. 1(d). The four islands divide the $Ni_{36}Fe_{64}$ wire into five F-sections connected in series by N-islands. For this 5-section sample, we have recorded FMR spectra (like the one in Fig. 1(e)) at different dc bias currents $I_{dc}$. Open black circles in Fig. 2(a) show the resulting FMR peak value of $V_\omega$ as a function of $I_{dc}$. The slope of this $V_\omega(I_{dc})$ dependence represents the extra resistance produced by FMR in the 5-section sample. We then cover one of the 5 F-sections by Ag paste, thus, effectively short-circuiting this F-section by the N-paste. This procedure results in a sample with only four active F-sections. For this 4-section sample, we again record FMR spectra at different dc bias currents $I_{dc}$. Open red circles in Fig. 2(a) show the resulting $V_\omega(I_{dc})$ with a smaller slope (compared to black circles), which represents a smaller extra resistance produced by FMR in the 4-section sample. We then continue to "switch off" (short-circuit) F-sections one by one, thus, reducing the number of active F-sections from five to zero, where the whole $Ni_{36}Fe_{64}$ wire becomes covered by Ag paste. Open green, blue, and magenta circles in Fig. 2(a) show $V_\omega(I_{dc})$ for samples with three, two, and one F-section, respectively. The sample with zero active F-sections (fully covered with Ag) did not show any $V_\omega$ induced by FMR (not shown), thus confirming an effective short-circuiting of $Ni_{36}Fe_{64}$ wire by Ag paste. Fig. 2(b) shows that the FMR-induced resistance $\Delta R_\omega$ [slope of $V_\omega(I_{dc})$ in Fig. 2(a)] increases with the number of active F-sections; here $S_N$ corresponds to a sample with N sections. We found that this extra resistance increases approximately linearly with the total length $L$ of active F-sections (see linear fit in Fig. 2(b)). This behavior correlates well with the previously observed [14] linear dependence of spin rectification on the length of a single F-wire. This single-wire geometry is equivalent to a series network where the wire length represents the number of F-elements connected in series. Next, we focus on a single $Ni_{36}Fe_{64}$ wire with variable length.

*Contact author: majibulbabu@utexas.edu

We have placed a 1.4-mm-long $Ni_{36}Fe_{64}$ wire into the circuit of Fig. 1(d) and reduced its length step by step by covering the wire partially with Ag paste. Here, the uncovered $Ni_{36}Fe_{64}$ wire represents the active F-wire whose length is controlled by the Ag coverage, which effectively short-circuits the rest of the wire. For the wire of a given active length: we measured FMR spectra at different dc bias currents $I_{dc}$ and determined the $V_\omega(I_{dc})$ dependence; the slope of the latter is the extra resistance $\Delta R_\omega$ induced by FMR. Open and filled circles in Fig. 3(a) show this extra resistance as a function of the active magnetic length for two independent sets of measurements. The open-circles data was taken from our previous measurements [14] and the filled circles are the new data, which show a good reproducibility of our results. The linear fit in Fig. 3(a) highlights the linear relationship between the FMR-induced resistance $\Delta R_\omega$ and the active length $L$ of F-wire. We have extended these measurements to extra-long $Ni_{36}Fe_{64}$ wires of up to 6 mm; the maximum length is limited by the magnet-gap size in our experimental setup. Fig. 3(b) shows that the extra resistance increases approximately linearly up to the longest wires used.

The observed linear dependence of the extra resistance produced by FMR on the length of magnetic conductor is analogous to the usual dependence of total network resistance in series circuits/single wires on the sum of individual resistors/conductor's length. We now turn our attention to F-elements connected in parallel to see if this analogy to classic electrical circuits holds for the parallel networks.

*Contact author: majibulbabu@utexas.edu

We now place our parallel network sample – six F wires ($Ni_{36}Fe_{64}$) connected in parallel – into the circuit of Fig. 1(d). For this 6-wire sample, we have recorded FMR spectra (like the one in Fig. 1(e)) at different dc bias currents $I_{dc}$. Open magenta circles in Fig. 4(a) show the resulting FMR peak value of $V_\omega$ as a function of $I_{dc}$. The slope of this $V_\omega(I_{dc})$ dependence represents the extra resistance $\Delta R_\omega$ produced by FMR in the 6-wire sample. We then start cutting out F-wires one by one, thus reducing the number $N$ of wires in our parallel-network sample. Open cyan, blue, green, red, and black circles in Fig. 4(a) show $V_\omega(I_{dc})$ for $N$ = 5, 4, 3, 2, and 1, respectively. Black squares in Fig. 4(b) show that the slope of these $V_\omega(I_{dc})$ dependencies, i.e., the extra resistance $\Delta R_\omega$ induced by FMR, decreases with increasing $N$. This is consistent with the classical circuit theory, where the equivalent resistance $R_{eq}$ of $N$ parallel resistors $R$ (wires) can be found by using reciprocal addition: $1/R_{eq}= 1/R+\ldots= N/R$, i.e., $R_{eq}= R/N$. However, the data in Fig. 4(b) shows a faster decrease of $\Delta R_\omega(N)$ compared with this expected $1/N$ dependence (see $1/N$ fit in Fig. 4(b)). This faster decrease of the FMR-induced resistance $\Delta R_\omega$ can be explained by a smaller rf current in individual F-wires of the parallel circuit and, therefore, its reduced effect on the wire resistance. The total rf current $j_{rf}$ is now split between $N$ wires that results in a smaller current ($\sim j_{rf}/N$) flowing through individual wires; this smaller current produces a lower rf magnetic field ($\sim h_{rf}/N$); the lower field produces smaller variations of the wire magnetization that, in turn, results in a smaller change of resistance $\Delta R_\omega$ of individual wires under FMR ($\sim j_{rf}/N$). In a parallel network, this extra $1/N$ factor adds up to the usual $1/N$ decrease of the equivalent resistance of N parallel resistors and results in $1/N \times 1/N = 1/N^2$ dependence of the FMR-induced resistance in parallel networks. This simple model shows much better agreement with our data (compare $1/N^2$ and $1/N$ fits in Fig. 4(b)), especially at higher $N$ numbers. The observed data deviations at low $N$s can be associated with a non-linear relationship between the rf excitation and resistance variations in the wire; for smaller $N$s, i.e., larger rf driving currents, such deviations are expected to be more pronounced.

## IV. Summary

We have studied experimentally the spin-rectification effect in parallel and series networks of ferromagnetic conductors. The networks were composed of $Ni_{36}Fe_{64}$ elements (wires) which, when subject to ferromagnetic resonance, develop a small spin-rectification photovoltage. We found that the effect of spin rectification can be modeled as an increase in resistance of the ferromagnetic elements and closely follows the usual dependence of total network resistance in series and parallel electric circuits: the total spin-rectification photovoltage in ferromagnetic wires connected in series is equal to the algebraic sum of contributions from individual wires, while in the parallel network one needs to add the reciprocals of the individual contributions. Our results support a simple way to control and manipulate the magnitude of spin-rectification photovoltage in multi-element spintronic devices.

*Contact author: majibulbabu@utexas.edu

*Contact author: majibulbabu@utexas.edu

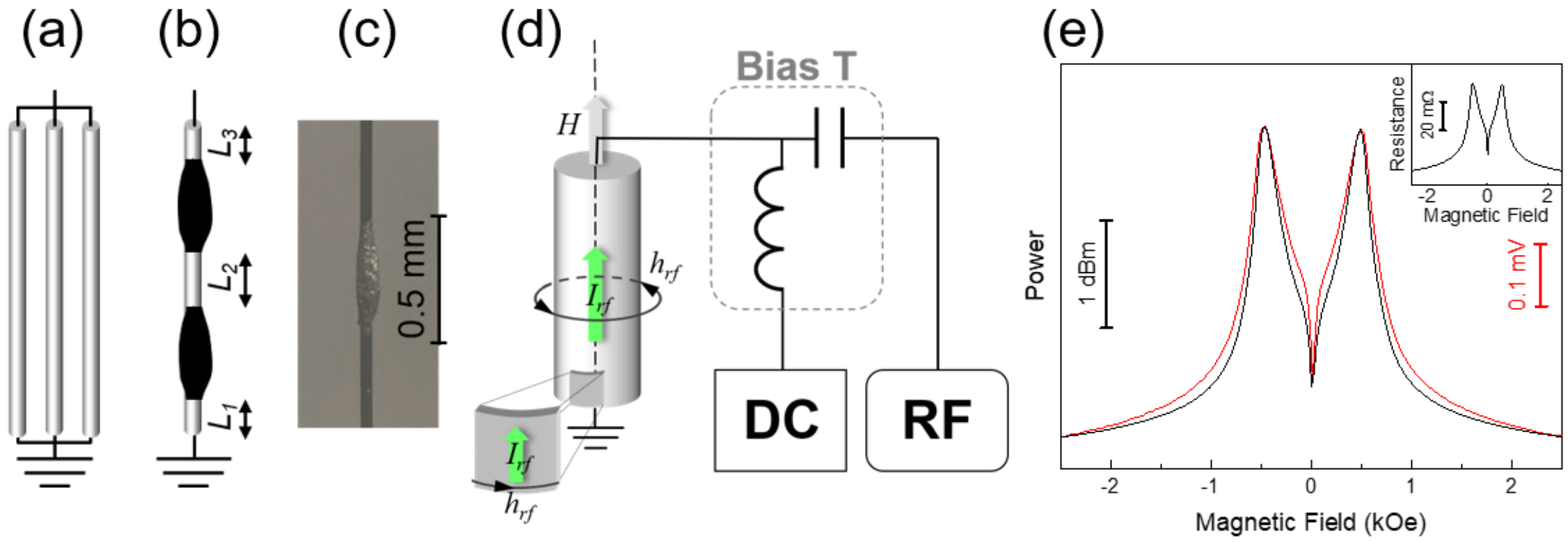

FIG 1. Experimental setup and device configurations. (a) Parallel F-network: up to $N = 6$ $Ni_{36}Fe_{64}$ wires were soldered between two common electrodes; here $N = 3$. (b) Series F-network: $Ni_{36}Fe_{64}$ wire soldered between the electrodes; a conducting Ag paste applied locally to create N-islands (black); here two N-islands divide the wire into three F sections $L_1$, $L_2$, and $L_3$; samples with up to four N-islands were used. (c) Microscope image of $Ni_{36}Fe_{64}$ wire with one N-island (lighter color in the middle). (d) Setup: dc and rf currents applied to a sample (here single $Ni_{36}Fe_{64}$ wire) via bias tee. (e) FMR spectra: absorption (black) and spin-rectification photovoltage (red) measured as a function of applied magnetic field. The insert shows the corresponding variation of the sample resistance. The peaks at ± 0.5 kOe show the extra resistance $\Delta R_\omega \approx 50$ mΩ induced by FMR. These measurements were performed on a single 1.4-mm-long $Ni_{36}Fe_{64}$ wire at $f$=7 GHz and $I_{dc}$ = 10 mA.

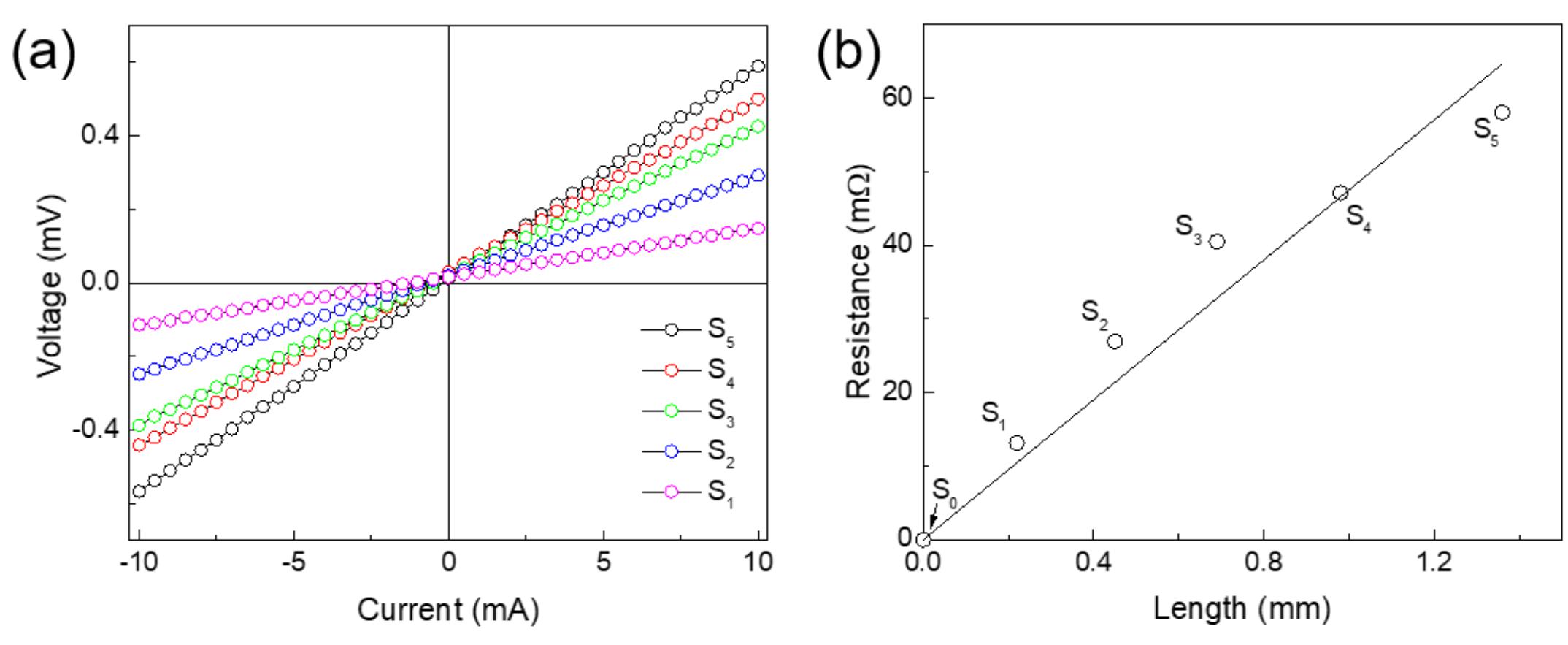

FIG 2. Spin rectification in series F-networks. (a) Open black, red, green, blue, and magenta circles show $V_\omega(I_{dc})$ for samples ($S_5$ - $S_1$) with five, four, three, two, and one F-section, respectively. The sample with zero active F-sections display no FMR and no $V_\omega$ (not shown). (b) Open symbols show the slope of $V_\omega(I_{dc})$ in (a), i.e., the FMR-induced resistance $\Delta R_\omega$, as a function of the total length $L$ of active F-sections. Black line is a linear fit (slope 48 mΩ/mm).

*Contact author: majibulbabu@utexas.edu

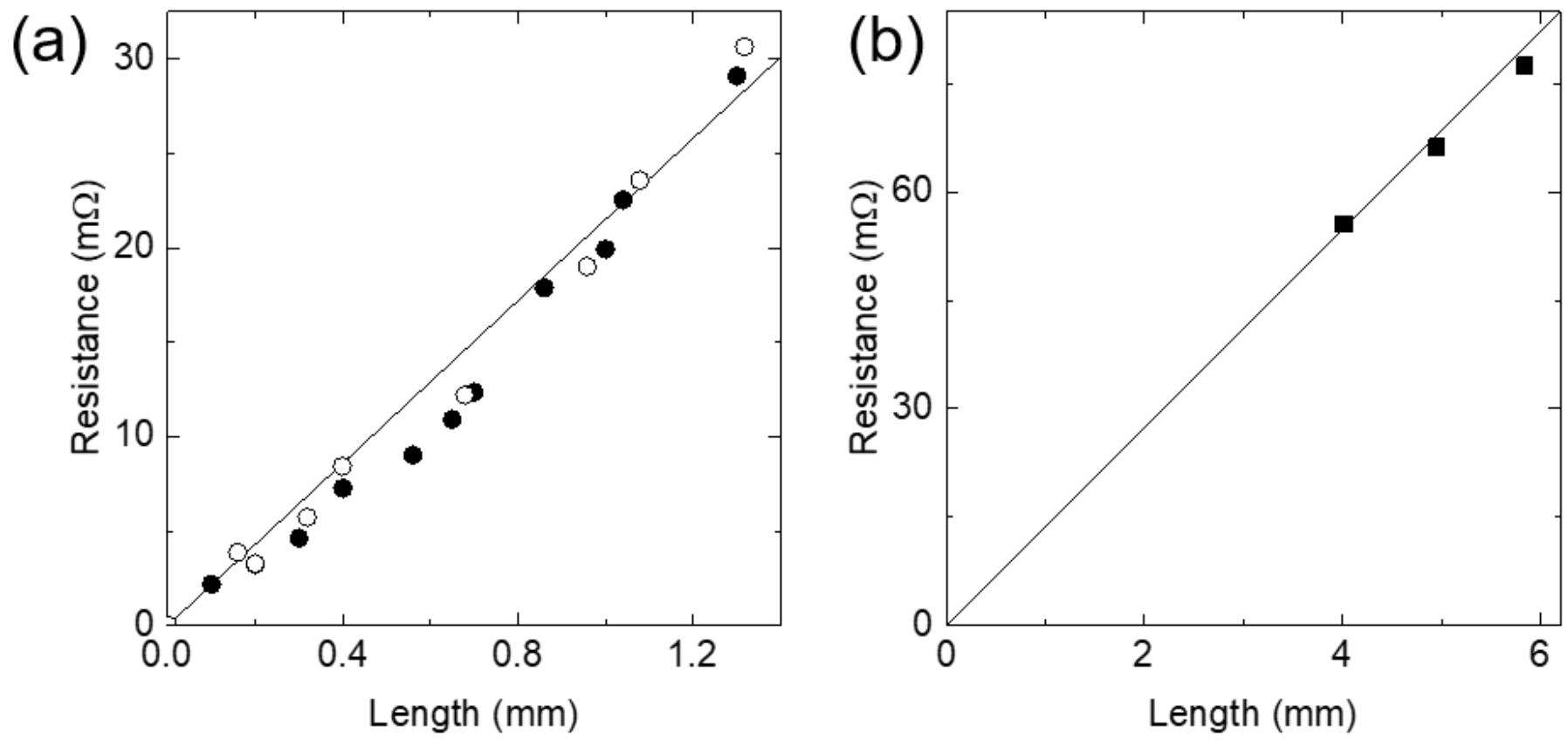

FIG 3. Spin-rectification vs active length in a single wire. (a) 1.4-mm-wire geometry: open and filled circles show the extra resistance $\Delta R_\omega$ induced by FMR as a function of the active magnetic length $L$ for two independent sets of measurements on different wires. Black line is a linear fit (slope 21 mΩ/mm). (b) 6-mm-wire geometry: black squares show $\Delta R_\omega(L)$. Black line is a linear fit (slope 14 mΩ/mm).

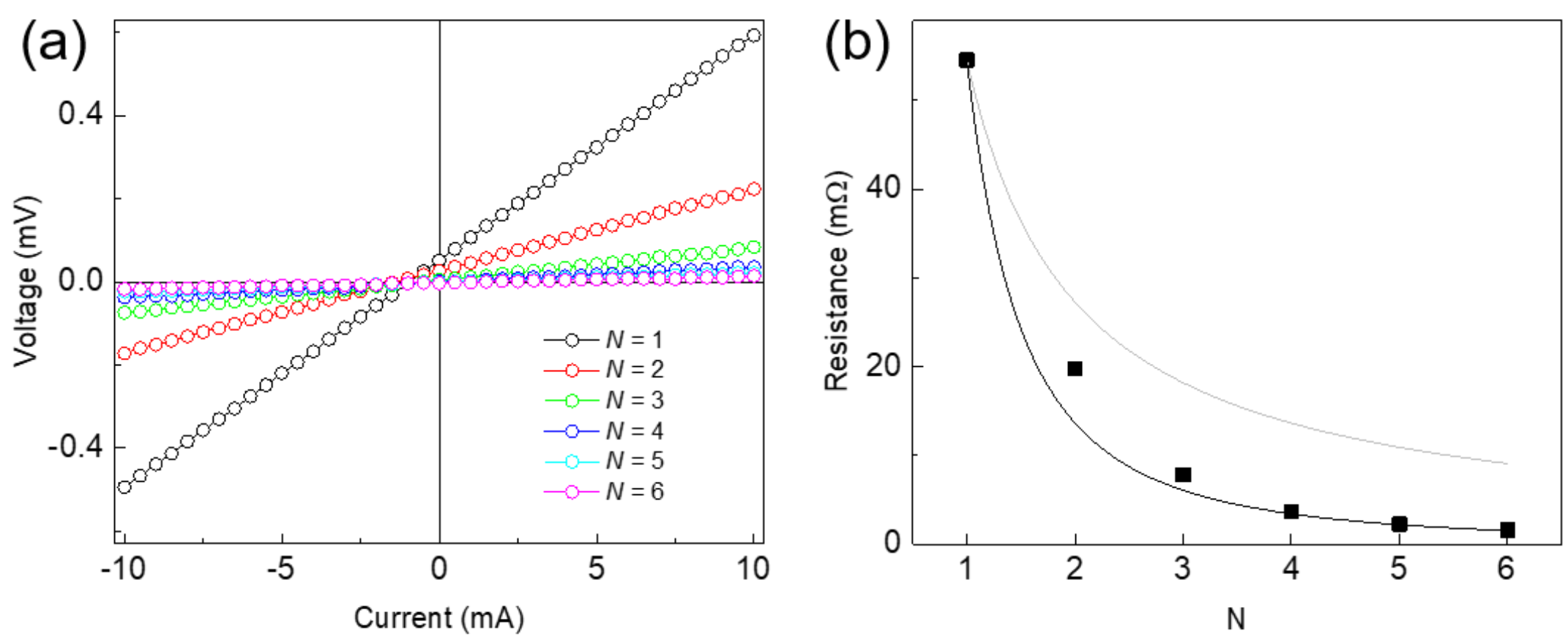

FIG 4. Spin rectification in parallel F-networks. (a) Open black, red, green, blue, cyan, and magenta circles show $V_\omega(I_{dc})$ for samples with $N$ = 1, 2, 3, 4, 5, and 6 parallel wires, respectively. (b) Black squares show the slope of $V_\omega(I_{dc})$ in (a), i.e., the extra resistance $\Delta R_\omega$ induced by FMR, versus $N$. Black (grey) curve is a $1/N^2$ ($1/N$) fit.

*Contact author: majibulbabu@utexas.edu